\documentclass{jpconf}
\usepackage{iopams}
\usepackage{graphicx}
\usepackage{subfigure}
\begin{document}

\title{MEM and CLEAN Imaging of VLBA Polarisation Observations of Compact Active Galactic Nuclei}

\author{Colm Coughlan}
\address{Department of Physics, University College Cork, Ireland}
\ead{colmcoughlanirl@gmail.com}

\author{Denise Gabuzda}
\address{Department of Physics, University College Cork, Ireland}
\ead{d.gabuzda@ucc.ie}

\begin{abstract}
 The Maximum Entropy Method (MEM) for the deconvolution of radio interferometry images is mathematically well based and presents a number of advantages over the
 usual CLEAN deconvolution, such as appreciably higher resolution. The application of MEM for polarisation imaging remains relatively little studied. CLEAN and MEM
 intensity and polarisation techniques are discussed in application to recently obtained 18cm VLBA polarisation data for a sample of Active Galactic Nuclei.
\end{abstract}

\section{Imaging Techniques in Radio Astronomy}

Consider a radio source which gives rise to a brightness distribution, $I(x,y)$, in the radio sky of the Earth. There exists a Fourier transform (FT) relationship
 between the true brightness distribution and the complex visibility function $V(u,v)$ measured by a VLBI array as a function of the baseline vector.

\begin{equation}
 V(u,v)=F.T.(I(x,y))
\end{equation}
\noindent
It is not a simple
 case of taking the inverse Fourier transform of the visbilities to return to the true brightness distribution as, due to the limited
 uv coverage offered by Earth rotation synthesis VLBI, most components of the visibility function are not measured. Mathematically this corresponds to multiplying
 the total visibility by a sampling function, $S(u,v)$, which eliminates most of the visibilities and leaves the observer with only $V_{sampled}(u,v)$.
 This means that, even before noise is taken into account, any attempt to recover the sky brightness distribution from the measured uv visibilites
 is a mathematically ill-posed problem.

\begin{equation}
 V_{sampled}(u,v)= V(u,v) S(u,v) + Noise
\end{equation}
\noindent
Deconvolution algorithms are used to attempt to interpolate the information contained in the measured visibilities to reconstruct
 the unmeasured visibilities and get as true an image of the sky brightness distribution as possible. There is no
 single ``best'' deconvolution algorithm, however the two most common are outlined below.

\subsection{The CLEAN Algorithm}

The CLEAN algorithm (H\"ogbom, 1974) is a standard technique used to deconvolve the measured visibilities and 
 attempt to recover the data from the
 unmeasured visibilities. First the measured visibilities are Fourier transformed to create a ``dirty'' map (containing only the
 information from the measured visibilities). This map is then described by a set of ``CLEAN'' components ($\delta$ functions).
 The set of ``CLEAN'' components is then convolved with the CLEAN beam (a Gaussian fit to the primary lobe of the point spread function)
 and the residual noise added in to give the final ``clean'' image. Images of Stokes I,Q and U parameters can be made independently using this technique.

\subsection{The Maximum Entropy Method (MEM)}

A second way to deconvolve the visibilities and recover some of the unmeasured data is to make a model intensity map of the source that satisfies specified mathematical criteria.
 This model can then be varied until the simulated visibilities agree with the actual data to within a specified limit, thus maximising the similarity of
 the model intensity and the true map. However, as the model is changed one cannot let the model visibilities converge
 to the actual measured visibilities, or the model would then simply yield the original ``dirty'' map. A regularising parameter is
 required to stop this convergence. In MEM, this parameter is the entropy of the image. Consider the following function:

\begin{equation}
 J=H(I_{m},P_{m})-\alpha \chi_{I}^{2}(V_{m},V_{d})-\beta \chi_{P}^{2}(V_{m},V_{d}) - conditions
\end{equation}
\noindent
where H is the entropy of the model map of the source and $\chi^{2}$ is a measure of the difference between the model (subscript m) and observed (subscript d)
 visibilities (there are two such terms, one for intensity, $\chi_{I}^{2}(V_{m},V_{d})$, and a second for polarisation, $\chi_{P}^{2}(V_{m},V_{d})$. $\alpha$ and $\beta$ are Lagrange
 parameters, and other conditions are also included which represent additional constraints, such as positivity of the Stokes I
 component. The optimal model of the source maximises the function J above. This results in a balance between entropy (representing
 noise, and the effect of unsampled visibilities) and fidelity to the observed data.\\

 Multiple forms of entropy can be used; common
 choices include Shannon entropy (\ref{shannon}), which is suitable for unpolarised emission, and the Gull and Skilling
 entropy (\ref{gull}), a generalised form of the Shannon entropy, suitable for polarised emission.

\begin{equation}
\label{shannon}
 H=-\sum_{k}I_{k} log(\frac{I_{k}}{B_{k}})
\end{equation}

\begin{equation}
\label{gull}
 H=-\sum_{k}I_{k} (log(\frac{I_{k}}{e B_{k}})+\frac{1+m_{i}}{2}log(\frac{1+m_{i}}{2})+\frac{1-m_{i}}{2}log(\frac{1-m_{i}}{2}))
\end{equation}
\noindent
 In these equations, the
 index $k$ represents summation over pixels, and $I_{k}$ and $m_{k}$ are the intensity and fractional polarisation, respectively, at pixel $k$.
 $B$ is a bias map, to which the solution defaults in the absence of data. See Holdaway \cite{holdaway} and Gull and Skilling \cite{gas}
 for more details about the MEM and different forms of polarisation entropy.

\section{CLEAN and MEM : A comparision}
\label{comparision}

Both CLEAN and MEM are widely used in image processing as deconvolution techniques, however each technique has specific
 strengths and weaknesses. CLEAN, while intuitive, does not have a firm mathematical footing, and it can be difficult to state
 the resolution of a CLEAN image exactly. MEM, while less intuitive, has a much better mathematical grounding, and has a resolution which can be
 demonstrated mathematically. As the deconvolution problem is inherantly ill-posed, neither technique is perfect - they cannot recreate the ``true''
 image, and different types of artefacts occur in each technique. This means that regardless of the imaging technique used, the images produced
 must be interpreted before any conclusions can be made about features that may be present in them.\\

In both techniques, the resolution obtained is inversely proportional to the size of the longest baseline in the observing array. Due
 to the lack of firm mathematical foundations for the CLEAN algorithm, a conservative estimation of its resolution is usually taken to be the full width at half
 maximum of the CLEAN beam, which is a Gaussian fit to the primary lobe of the point spread function. Some information is also present on smaller scales,
 but can be difficult to interpret. MEM's resolution can be shown to be

\begin{equation}
\label{eqn:res}
x_{min}=\frac{1}{4 u_{max}}
\end{equation}
\noindent
where $x_{min}$ is the resolution in radians, and $u_{max}$ is the length of the longest baseline in wavelengths \cite{holdaway}. This resolution is approximately four
 times smaller than the conservative estimate of the CLEAN algorithm.

\section{Application of polarisation MEM to VLBI data}

The standard radio astronomy software package Astronomy Image Processing System (AIPS) includes a task to conduct MEM deconvolution of intensity (Stokes I),
 but not polarisation (Stokes Q and U) images (see Cornwell and Evans \cite{cae} for details). The relation between the Stokes Q and U parameters and the polarised intensity and polarisation angle of a map is
 shown by equations (\ref{poli}) and (\ref{pola}).

\begin{equation}
\label{poli}
P=\sqrt{Q^{2}+U^{2}}
\end{equation}

\begin{equation}
\label{pola}
\chi=\frac{1}{2}ArcTan(\frac{U}{Q})
\end{equation}
\noindent
We are in the process of investigating a number of ways of implementing fully polarised MEM to VLBI polarisation data.
 We have in the meantime produced a number of ``proof of concept'' MEM polarisation maps in AIPS by devising an algorithm to work around the limitations imposed by the AIPS
 MEM task (primarily the non-negativity requirement - whereas Stokes I must be positive, Stokes Q and U can be either positive or negative). A major limitation of the work-around devised was the assumption
 that the location of positive and negative pixels in the Q and U maps remained the same as in the original maps (only their amplitude was allowed to vary).

\section{Preliminary MEM Intensity and Polarisation VLBA Images}

The prelimary images shown were obtained with the Very Long Baseline Array (see Coughlan et al \cite{cc}). Both the CLEAN and MEM images were made from the same uv data, which had been calibrated using standard methods in AIPS. The
 CLEAN map was convolved with the CLEAN beam and the MEM map convolved with a Gaussian beam corresponding to the resolution indicated by (\ref{eqn:res}).
 In all cases, the MEM images lost some of the extended emission visible in the CLEAN images. However, the increased resolution of the MEM images provides fuller information
 about inner jet structure and morphology. The superimposed sticks indicate the local polarisation angles. In all cases there is good agreement between the MEM and CLEAN
 polarisation angles.

\begin{figure}
\begin{center}
\includegraphics[scale=0.55]{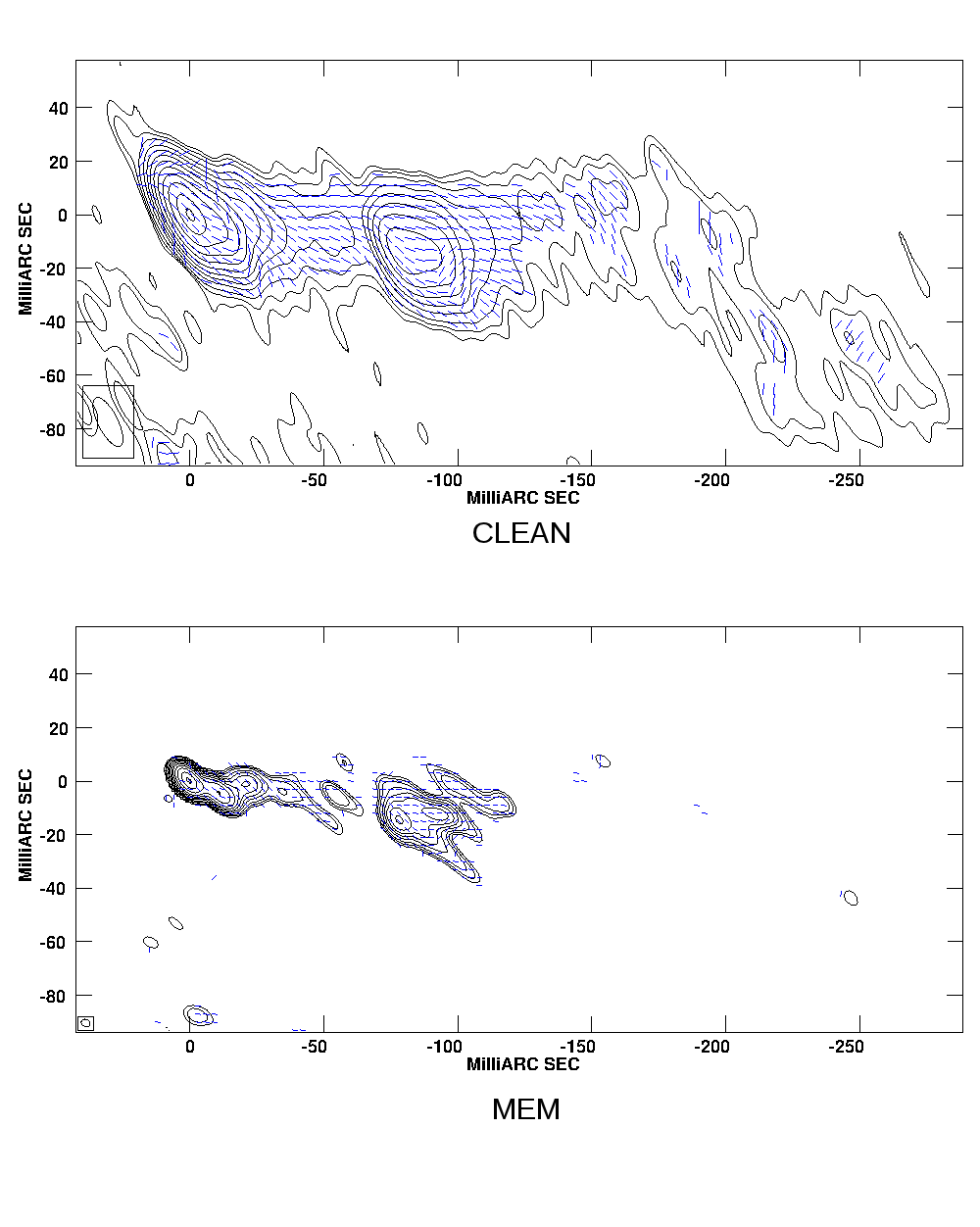}
\caption{CLEAN and MEM images of J0433+0521 (3C120) at 1.36 GHz.}
\label{fig:J0433}
\end{center}
\end{figure}

\begin{figure}
\begin{center}
\includegraphics[scale=0.55]{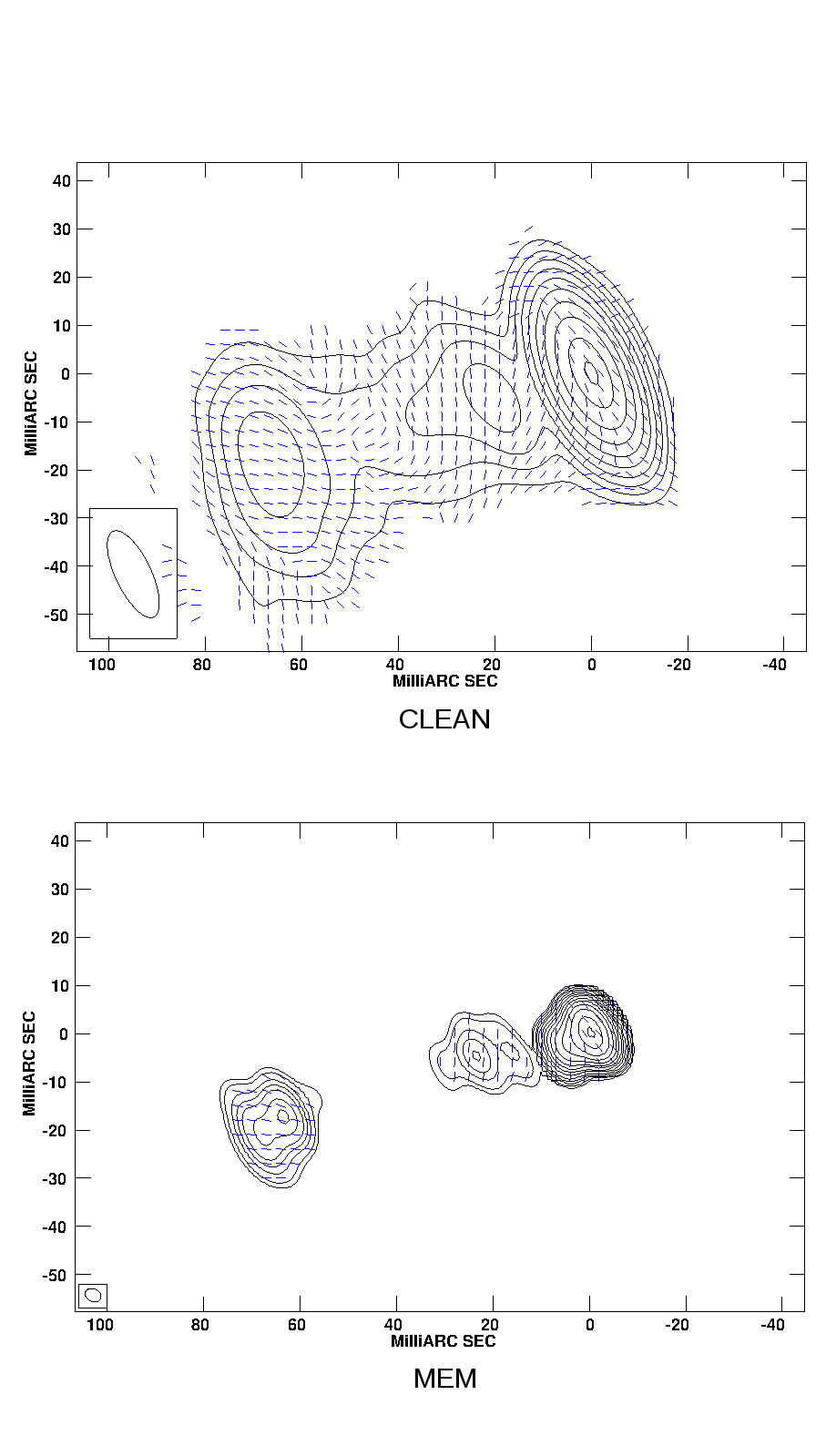}
\caption{CLEAN and MEM images of J0750+1231 at 1.36 GHz.}
\label{fig:J0750}
\end{center}
\end{figure}

\begin{figure}
\begin{center}
\includegraphics[trim = 0mm 22mm 0mm 12mm, clip,scale=0.47]{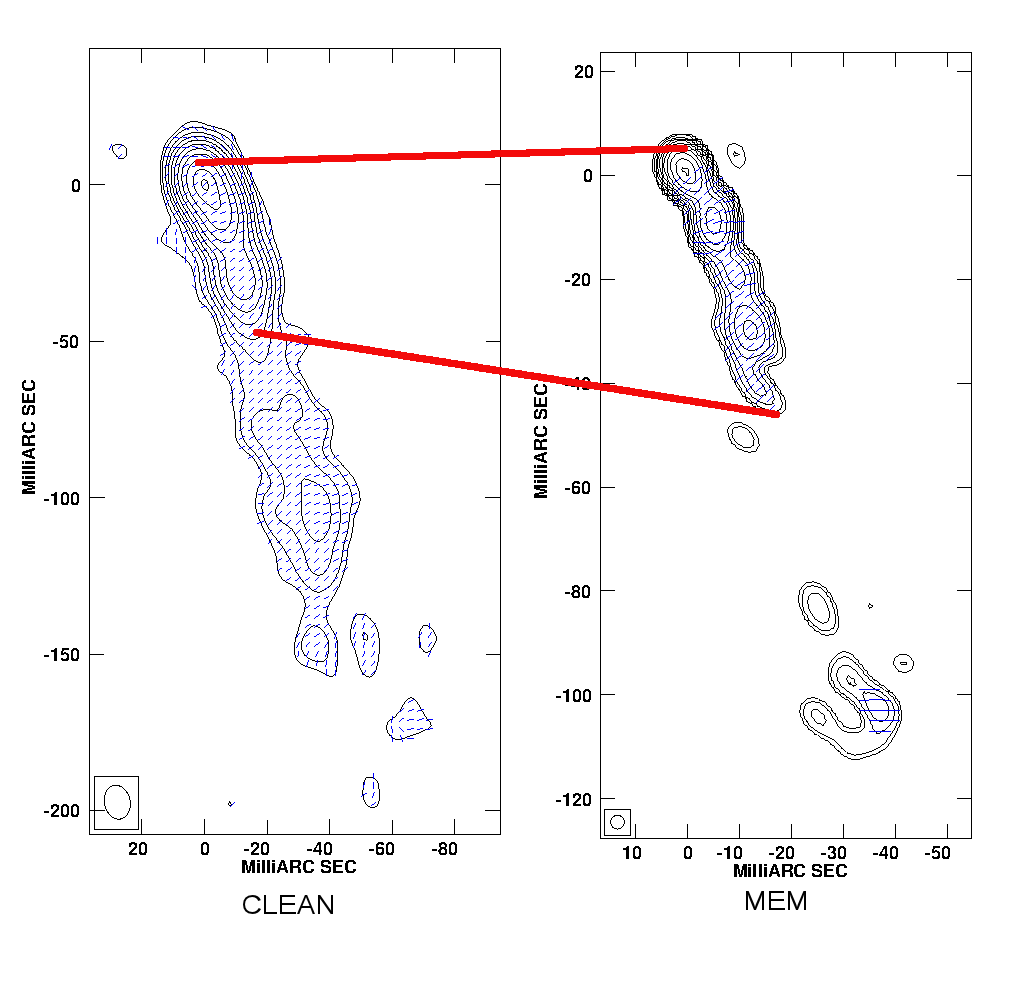}
\end{center}
\caption{CLEAN and MEM images of J0841+7053 at 1.36 GHz.}
\label{fig:J0841}
\end{figure}

\section{Conclusion}

Both CLEAN and MEM offer unique and complementery perspectives on the same uv visibility data. MEM appears less sensitive to low intensity emission, but provides higher resolution, allowing jet direction
 and morphology to be studied in more detail. There is good agreement between the CLEAN and approximate MEM polarisation
maps we have constructed, suggesting that the approximations made in the work-around used
 to produce polarised MEM maps in AIPS have not adversely affected the polarisation. However, in order to generate a truely reliable MEM polarisation map a form of polarised entropy must be used
 and we are in the process of implementing the entropy (\ref{gull}) for this purpose.
 The synergy between CLEAN and MEM in studying polarised emission should then lead to a clearer picture of the polarisation intensity and direction along the jet.
 We also plan to investigate the construction of MEM spectral-index and Faraday-rotation maps based on multi-frequency intensity and polarisation VLBI maps.

\section*{Acknowlegements}

This work was supported by the Irish Research Council for Science, Engineering and Technology (IRCSET).

\end{document}